\documentstyle[amsfonts,aps,psfig,epsfig]{revtex}

\newcommand{\be}{\begin{equation}}
\newcommand{\ee}{\end{equation}}
\newcommand{\br}{\begin{eqnarray}}
\newcommand{\er}{\end{eqnarray}}

\newcommand{\bd}{\begin{displaymath}}
\newcommand{\ed}{\end{displaymath}}

\newcommand{\bfig}{\begin{figure}}
\newcommand{\efig}{\end{figure}}

\def\3cdot{\cdot \cdot \cdot}

\def\om0{\omega _0}
\def\Om0{\Omega _0}

\def\text#1{{\rm{#1}}}

\def\->{\rightarrow}
\def\=>{\Rightarrow}
\def\-->{\longrightarrow}
\def\==>{\Longrightarrow}

\def\pr{^\prime}
\def\pr2{^{\prime\prime}}

\def\bfig{\begin{figure}}
\def\efig{\end{figure}}

\begin{document}
\title{The Wigner function associated to the Rogers-Szeg\"{o} polynomials}
\author{$^{1}$D. Galetti\thanks{%
E-mail:galetti@ift.unesp.br}, $^{2}$S. S. Mizrahi\thanks{%
E-mail: salomon@df.ufscar.br}, $^{3}$M. Ruzzi\thanks{%
E-mail:mruzzi@fsc.ufsc.br}}
\address{$^1$Instituto de F\'{i}sica Te\'{o}rica (IFT), Universidade\\
Estadual Paulista (UNESP)\\
Rua Pamplona 145, 01405-900, S\~{a}o Paulo, SP, Brazil. \\
$^2$Departamento de F\'{\i}sica, CCET, Universidade Federal de\\
S\~{a}o Carlos,\\
Via Washington Luiz km 235, 13565-905, S\~ao Carlos, SP, Brazil.\\
$^{3}$Departamento de F\'{i}sica, Universidade Federal de Santa Catarina,\\
88040-900,\\
Florian\'{o}polis, SC, Brazil.}
\date{\today}
\maketitle

\begin{abstract}
We show here that besides the well known Hermite polynomials, the $q$%
-deformed harmonic oscillator algebra admits another function space
associated to a particular family of $q$-polynomials, namely the Rogers-Szeg%
\"{o} polynomials. Their main properties are presented, the associated
Wigner function is calculated and its properties are discussed. It is shown
that the angle probability density obtained from the Wigner function is a
well-behaved function defined in the interval $-\pi \leq \theta <\pi $,
while the action probability only assumes integer values $m\geq 0$. It is
emphasized the fact that the width of the angle probability density is
governed by the free parameter $q$ characterizing the polynomial.
\end{abstract}

\pacs{02.20.Uw, 02.30.Gp, 03.65.-w}


%

\section{Introduction}

%
Deformed algebras, quantum groups and quantum spaces have been thoroughly
investigated since their introduction in physics \cite
{kulish,fadeev,jimbo,drinfeld,manin} and non-trivial commutation relations
have been proposed and analyzed by many authors \cite
{iwata,arcoon,kurysh,cigler,jan81}. Following this trend, the $q$-deformed
harmonic oscillator was introduced in the articles \cite
{biedenharn,macfarlane} and has been the recipient of further investigation 
\cite
{kudam,atsus,sokat,fiv91,celeg91,chasri91,jan91,chiu92,ell92,manmen,alsol,camp94,shan94,ck96,katq96,park96,boda99,isar02,manko1,artbir,gruv,chak91,arik92,chung,3russos,manko}%
. Many kinds of 3-generator algebras $\left\{ A,A^{\dagger },N\right\} $
were proposed, whose commutation relations take the general form 
\begin{equation}
AA^{\dagger }-f_{1}(q,\alpha ,\beta ,\gamma ,...)A^{\dagger
}A=f_{2}(q,\alpha ,\beta ,\gamma ,...;N)  \label{commut1}
\end{equation}
\begin{equation}
\left[ N,A\right] =-A,\qquad \left[ N,A^{\dagger }\right] =-A^{\dagger },
\label{commut2}
\end{equation}
where $f_{1}(q,\alpha ,\beta ...)$ is a $c$-number function of the
parameters $q,\alpha ,\beta ,\gamma ,...$ (real or complex) and $%
f_{2}(q,\alpha ,\beta ...;N)$ is an operator function since it also depends
on $N$. Many realizations or representations are possible for $\left\{
A,A^{\dagger },N\right\} $, however, when $f_{1}(q,\alpha ,\beta ,...)=1$
and $f_{2}(q,\alpha ,\beta ,...;N)=1$, the linear harmonic oscillator
algebra is recovered. Each realization allows a specific space of functions
on which the operators act; different spaces show quite different properties
and different physics. For instance, setting $f_{1}(q,\alpha ,\beta ,...)=1$
and $f_{2}(q,\alpha ,\beta ,\gamma ,...;N)=q^{N}$, two different
realizations for the space of functions are possible: (a) the ${\Bbb L}^{2}$
space of Hermite polynomials, with a Gaussian weight \cite
{jan81,manko,micamdod}, that have the line ${\Bbb R=}\left\{ x\in \left(
-\infty ,\infty \right) \right\} $ as the domain; and the less familiar (b)
Rogers-Szeg\"{o} polynomials which have the Jacobi $\vartheta _{3}$-function
as weight function, the independent variable is an ``angle'', $\phi $, with
compact domain ${\Bbb S=}\left\{ \phi \in \lbrack 0,2\pi )\right\} $.

Though the properties of the Wigner function associated to the Hermite
polynomial are quite known, the same can not be said about the
Rogers-Szeg\"{o} polynomials, so it is the aim of the present paper to study
their properties, to derive the associated Wigner function and the angle
distribution function.

We remind that the Weyl-Wigner transformation associated to the well-known
translational degree of freedom for Cartesian variables and moments of a
particle has long been established and widely discussed in the literature 
\cite{degroot,carruthers,balazs,hillery,kim,ozorio}. On the other hand, in
what refers to the Weyl-Wigner transformation, the rotational degree of
freedom has been scarcely touched upon. In this connection, the treatment of
this case was directly inferred from the previous one by means of symmetry
arguments \cite{berry,mukunda}, by the continuous limit of
finite-dimensional Weyl-Wigner mappings \cite{galetti}, or by the
implementation of the appropriate kinematics relations \cite{bizarro}. It is
clear in all these cases that one is dealing with functions of angular
variable that have period $2\pi $ and the measure of this function space is
simply the unity.

In this contribution we intend first of all to briefly present a family of $%
q-$ polynomials that also have period $2\pi $, but are orthonormalized on
the circle with respect to a measure function $\mu \left( \theta ;q\right) $
that is the Jacobi $\vartheta _{3}$- function, namely, the Rogers-Szeg\"{o}
polynomials \cite{szego,szego2,carlitz1,carlitz,andrews}. This set of
polynomials has been shown to be associated to a realization of the $q-$
deformed harmonic oscillator \cite{macfarlane,galetti2}, and are then the
functions describing the states of that system for the deformation parameter
ranging from $0$ to $1$. Therefore, these functions are characterized by a
discrete variable $n$ and a continuous angle variable $\theta $, defined on
the circle, besides depending on the deformation parameter $q$. We then
write the Weyl-Wigner transformation for this class of $q-$ polynomials from
which we also extract the angle and action probability distribution
functions. The simplest cases of the ground state and first state are
directly obtained and discussed.

In section II we present a brief review of definitions and relations that
are relevant in the present context for the study of the Rogers-Szeg\"{o}
polynomials which are also presented. The Wigner function for these
polynomials are calculated in section III, where the probability
distributions for $n$ and $\theta $ are also presented. Finally, section IV
is devoted to the summary and a discussion of what has been obtained. %

\section{q-series, Rogers-Szeg\"{o} polynomials and q-deformed algebra}

%

\subsection{Brief review of some results from q-series}

%
In order to set the stage for the introduction of the $q$-polynomials we are
interested in, let us first introduce some basic concepts\cite
{andrews,gasper}. The basic notation will be, for $\left| q\right| <1$, 
\[
\left( x\right) _{n+1}\equiv \left( x;q\right) _{n+1}\equiv \left(
1-x\right) \left( 1-xq\right) \left( 1-xq^{2}\right) \ldots \left(
1-xq^{n}\right) 
\]
such that 
\[
\left( x\right) _{\infty }\equiv \left( x;q\right) _{\infty }\equiv
\lim_{n\rightarrow \infty }\;\left( x;q\right) _{n}; 
\]
and 
\[
\left( x\right) _{0}\equiv 1. 
\]
The particular case $x=q$ should be noted 
\[
\left( q\right) _{n+1}\equiv \left( q;q\right) _{n+1}\equiv \left(
1-q\right) \left( 1-q^{2}\right) \left( 1-q^{3}\right) \ldots \left(
1-q^{n+1}\right) . 
\]
Furthermore, we can recognize that, for any real $n$, we can write 
\[
\left( x\right) _{n}=\frac{\left( x\right) _{\infty }}{\left( xq^{n}\right)
_{\infty }}=\frac{\left( 1-x\right) \left( 1-xq\right) \left(
1-xq^{2}\right) \ldots \left( 1-xq^{n+1}\right) \ldots }{\left(
1-xq^{n}\right) \left( 1-xq^{n+1}\right) \ldots } 
\]
\begin{equation}
=\left( 1-x\right) \left( 1-xq\right) \left( 1-xq^{2}\right) \ldots \left(
1-xq^{n-1}\right) .  \label{II1}
\end{equation}
The $q$-binomial is defined as 
\begin{equation}
{%
{n  \atopwithdelims[] j}%
} =\frac{\left( 1-q\right) \left( 1-q^{2}\right) ...\left( 1-q^{n}\right) }{%
\left( 1-q\right) ...\left( 1-q^{j}\right) \left( 1-q\right) ...\left(
1-q^{n-j}\right) .}=\frac{\left( q\right) _{n}}{\left( q\right) _{j}\left(
q\right) _{n-j}}  \label{II2}
\end{equation}
for $j$ and $n$ integers, with $0\leq j\leq n$ and $(0)_{n}=1$, and has the
following properties 
\begin{equation}
{%
{n  \atopwithdelims[] 0}%
} = {%
{ n  \atopwithdelims[] n}%
} =1,  \label{II3}
\end{equation}
\begin{equation}
{%
{n  \atopwithdelims[] j}%
} ={%
{n  \atopwithdelims[] n-j}%
} ,  \label{II4}
\end{equation}
\begin{equation}
\lim_{q\rightarrow 1}{%
{n  \atopwithdelims[] j}%
} ={%
{ n  \choose j}%
} =\frac{n!}{j!\left( n-j\right) !}.  \label{II7}
\end{equation}

The so-called $q$-number can be directly realized from the definition (\ref
{II2}), 
\[
{%
{n  \atopwithdelims[] 1}%
} =\frac{1-q^{n}}{1-q}\equiv \left[ n\right]. 
\]
Cauchy theorem plays an essential role in this context and states that, for $%
\left| t\right| <1$, the following equality holds 
\[
\prod_{n=0}^{\infty }\frac{1-xtq^{n}}{1-tq^{n}}=\sum_{n=0}^{\infty }\frac{%
\left( x;q\right) _{n}}{\left( q;q\right) _{n}}t^{n} 
\]
from which follows the important result 
\[
\left( x;q\right) _{n}=\prod_{s=0}^{n-1}\left( 1-q^{s}x\right)
=\sum_{j=0}^{n}\left( -1\right) ^{j}{\ {%
{n  \atopwithdelims[] j}%
} }q^{j\left( j-1\right) /2}x^{j}. 
\]
%

\subsection{Rogers-Szeg\"{o} polynomials}

%
Let us introduce the Rogers-Szeg\"{o} polynomials in the general form
through their definition \cite{szego,carlitz1,carlitz} 
\begin{equation}
H_{n}\left( y \right) \equiv H_{n}\left( y;q\right) =\sum_{r=0}^{n} {%
{n  \atopwithdelims[] r}%
} \;y^{r}.  \label{III1}
\end{equation}
Among the several properties they satisfy, we call attention on the
three-term recurrence relation, 
\begin{equation}
H_{n+1}\left( y;q\right) =\left( 1+y\right) H_{n}\left( y;q\right) -\left(
1-q^{n}\right) yH_{n-1}\left( y;q\right) ,  \label{recur1}
\end{equation}
and the $q$-differentiation relation, 
\begin{equation}
D_{q}H_{n}\left( y;q\right) =\frac{H_{n}\left( y;q\right) -H_{n}\left(
yq;q\right) }{y\left( 1-q\right) }=\left[ n\right] H_{n-1}\left( y;q\right) ,
\label{deriv1}
\end{equation}
where $D_{q}$ is known as Jackson' s $q$-derivative, which goes to the usual
derivative in the limit $q\rightarrow 0$.

From the definition (\ref{III1}) and properties (\ref{II3}) and (\ref{II4}),
the first two polynomials are 
\begin{equation}
H_{0}\left( y;q\right) =1,  \label{III2}
\end{equation}
\begin{equation}
H_{1}\left( y;q\right) =1+y.  \label{III3}
\end{equation}
The other ones can be obtained through the use of the recurrence relation (%
\ref{recur1}). In the limit $q\rightarrow 1$ we have 
\[
\lim_{q\rightarrow 1}H_{n}\left( y;q\right) =\sum_{r=0}^{n} {%
{n  \choose r}%
} \;y^{r}=\left( 1+y \right) ^{n}. 
\]

Another important property of the Rogers-Szeg\"{o} polynomials is their
orthogonality on the circle when the Jacobi $\vartheta _{3}\left( y;q\right) 
$ function is taken as the measure function\cite{szego2}. In order to
explicitly verify this, we should perform a proper choice for the variable $%
y $, $y=$ $-q^{-1/2}e^{i\varphi }$, such that 
\begin{equation}
H_{n}\left( y;q\right) =H_{n}\left( -q^{-1/2}e^{i\varphi };q\right) .
\label{III5}
\end{equation}
In this form, the orthonormalization integral is written as 
\[
I_{mn}\left( q\right) =\int_{-\pi }^{\pi }H_{m}\left( -q^{-1/2}e^{i\varphi
};q\right) H_{n}\left( -q^{-1/2}e^{-i\varphi };q\right) \vartheta _{3}\left(
\varphi ;q\right) \frac{d\varphi }{2\pi } 
\]
with the explicit form 
\begin{equation}
\vartheta _{3}\left( \varphi ;q\right) =\sum_{m=-\infty }^{\infty
}q^{m^{2}/2\;}e^{im\varphi }=\sum_{m=-\infty }^{\infty }e^{-\mu
m^{2}+im\varphi },  \label{III6}
\end{equation}
with $\mu =-\left( \ln q\right) /2$, which is the measure function \cite
{whittaker}. Using the definition of the Rogers-Szeg\"{o} polynomials, Eq. (%
\ref{III1}), we see that 
\begin{equation}
I_{mn}\left( q\right) =\sum_{r=0}^{m}\sum_{s=0}^{n}\left( -1\right) ^{r+s}{%
{m  \atopwithdelims[] r}%
}{%
{ n  \atopwithdelims[] s}%
} q^{r\left( r-1\right) /2}q^{s\left( s-1\right) /2}q^{-rs};  \label{III6a}
\end{equation}
this is a result discussed by Carlitz\cite{carlitz} (see Appendix for the
proof), and shown to give 
\begin{equation}
I_{mn}\left( q\right) =\frac{(q,q)_{n}}{q^{n}}\;\delta _{m,n}.  \label{III7}
\end{equation}
From that result we get what is sometimes known as the Rogers-Szeg\"{o}
functions 
\begin{equation}
R_{n}\left( \varphi ;q\right) =\frac{q^{n/2}}{\left[ \left( q,q\right) _{n}%
\right] ^{1/2}}\;H_{n}\left( -q^{-1/2}e^{i\varphi };q\right) .  \label{III8}
\end{equation}

It is worth noting that the Jacobi $\vartheta _{3}\left( \varphi ;q\right) $
function is associated to a sum of Gaussians on the circle. To see this, let
us recall the basic relation obeyed by that function, namely, in a general
form\cite{whittaker,bellman} 
\begin{equation}
\sum_{n=-\infty }^{\infty }\exp \left[ -\alpha \left( y+n\right) ^{2}\right]
=\sqrt{\frac{\pi }{\alpha }}\sum_{k=-\infty }^{\infty }\exp \left( -\pi ^{2} 
\frac{k^{2}}{\alpha }+2\pi iky\right) .  \label{III9}
\end{equation}
Adapting to our case, we take $q=\exp \left( -2\mu \right) $, and using Eq.
( \ref{III6}), we have 
\begin{equation}
\sum_{m=-\infty }^{\infty }\exp \left( -\mu m^{2}-im\varphi \right) =\sqrt{%
\frac{\pi }{\mu }}\sum_{n=-\infty }^{\infty }\exp \left[ -\frac{1}{4\mu }
\left( \varphi -2\pi n\right) ^{2}\right] ;  \label{III10}
\end{equation}
the RHS of this relation shows the measure function as a sum of Gaussians of
the angle variable $\varphi $.

From a different point of view, the Jacobi $\vartheta _{3}$-function has
been proposed as a valuable function to describe particular limiting
situations in quantum optics\cite{russo}, and also as a coherent state for a
particle on a circle where the angular variable plays now an essential role 
\cite{gonz,kowa}. In this case, the algebra is given in terms of the angular
momentum and an unitary operator so that the commutation relation is 
\[
\left[ J,U\right] =U, 
\]
being U a unitary operator associated to angular momentum shifts. Such a
commutation relation has been discussed long ago in the literature \cite
{carr,suss}, and was also obtained as the limiting case in finite
dimensional phase space representation of quantum mechanics \cite{galetti}.
It is worth noticing that the Jacobi $\vartheta_{3}$-function with integer
argument was also proposed as a coherent state for the case of any
finite-dimensional degrees of freedom \cite{gama}, since in these cases the
eigenvalue problem associated to the discrete Fourier matrix in the discrete
basis\cite{mehta} gives a solution which is directly expressed in terms of
that Jacobi's function. In this sense we see that the Jacobi $\vartheta_{3}$
-function plays a wider role in connection with coherent states, and, in
particular, with the rotational or action-angle degrees of freedom.

Here, instead, the Jacobi $\vartheta _{3}$-function will be considered the
measure function for the Rogers-Szeg\"{o} polynomials, in the same way as
the Gaussian is the measure function for the standard Hermite polynomials
associated to the one-dimensional harmonic oscillator.

The angle probability distribution constructed out of the $\vartheta_{3}$
-function and the Rogers-Szeg\"{o} polynomials is in fact a good candidate
for describing action-angle degrees of freedom, since if we consider the
Rogers-Szeg\"{o} functions, the angle probability distribution associated to
these states is immediately recognized as 
\[
\Omega ^{\left( n\right) }\left( \varphi ;q\right) d\varphi =\left|
R_{n}\left( \varphi ;q\right) \right| ^{2}\vartheta _{3}\left( \varphi
;q\right) \frac{d\varphi }{2\pi }. 
\]
In the same form we have for the harmonic oscillator functions on the line 
\[
\Phi ^{\left( n\right) }\left( x;b\right) dx=\left| H_{n}\left( x;b\right)
\right| ^{2}\exp \left( -\frac{x^{2}}{b^{2}}\right) dx, 
\]
where $b$ in the second case stands for the harmonic oscillator width and $%
H_{n}\left( x;b\right) $ is the standard Hermite polynomial.

As a further remark we observe that with the particular choice of the
deformation parameter, namely, $q=\exp \left( -2\mu \right) $, associated to
the range $0\leq q\leq 1$, we have $0\leq \mu \leq \infty $. Obviously in
the limit $q\rightarrow 1\;$($\mu \rightarrow 0$) we verify that the $%
\vartheta _{3}$-function is simply the usual Fourier series while 
\[
\lim_{q\rightarrow 1}H_{n}\left( y;q\right) =\sum_{r=0}^{n} 
{%
{n  \choose r}%
} \;y^{r}=\left( 1+y\right) ^{n}. 
\]
%

\subsection{ The $q$-deformed algebra}

%
Now, from Eqs. (\ref{recur1}) and (\ref{deriv1}) we show that the
Rogers-Szeg\"{o} polynomials make the function space realization of the
one-parameter, three-element $q$-deformed algebra, 
\begin{equation}
\left[ A,A^{\dagger }\right] =q^{N};\qquad \left[ N,A^{\dagger }\right]
=A^{\dagger },\qquad \left[ N,A\right] =-A  \label{commut3}
\end{equation}
and 
\[
A^{\dagger }A=\frac{1-q^{N}}{1-q}=\left[ N\right] , 
\]
such that $\lim_{q\rightarrow 1}A^{\dagger }A=N.$ Under certain requirements 
\cite{feinsilver} the commutation relation $\left[ A,A^{\dagger }\right]
=q^{N}$ can be shown to be cast in the form \cite{arcoon,atsus} 
\[
AA^{\dagger }-qA^{\dagger }A=1. 
\]
Also, it is immediate to verify that for any analytic function $f(x)$ one
has the relations, 
\[
Af(N)=f(N+1)A,\qquad A^{\dagger }f(N)=f(N-1)A^{\dagger }. 
\]
Now setting a representation to the operators $A$ and $A^{\dagger }$\cite
{galetti2}, 
\[
A=D_{q},\qquad A^{\dagger }=\left( 1+y\right) -\left( 1-q\right) yD_{q}, 
\]
one verifies that 
\[
A^{\dagger }H_{n}\left( y;q\right) =H_{n+1}\left( y;q\right) ,\qquad
AH_{n}\left( y;q\right) =\left[ n\right] H_{n-1}\left( y;q\right) 
\]
and 
\[
NH_{n}\left( y;q\right) =nH_{n}\left( y;q\right) ,\qquad A^{\dagger
}AH_{n}\left( y;q\right) =\left[ n\right] H_{n}\left( y;q\right) . 
\]

It must be remarked that the Rogers-Szeg\"{o} polynomials have also appeared
in a different way in the treatment of the $q-$deformed harmonic oscillator
by MacFarlane\cite{macfarlane}: Doing a nonlinear transformation on $A$, 
\[
B=\left( q-q^{-1}\right) ^{1/2}Aq^{-N/2}=\left( 1-q^{-2}\right)
^{1/2}q^{-N/2}A, 
\]
the new operators obey the $q-$commutation relation 
\begin{equation}
q^{2}BB^{\dagger }-B^{\dagger }B=q^{2}-1,  \label{III11}
\end{equation}
having the differential representation, 
\begin{equation}
B=\exp \left( -i\varphi \right) -\exp \left( -\frac{i\varphi }{2}\right)
\exp \left( -4i\mu \frac{\partial }{\partial \varphi }\right)  \label{III12}
\end{equation}
and 
\begin{equation}
B^{+}=\exp \left( i\varphi \right) -\exp \left( 4i\mu \frac{\partial }{
\partial \varphi }\right) \exp \left( \frac{i\varphi }{2}\right) ,
\label{III13}
\end{equation}
where $q=\exp \left( -2\mu \right) $.

On this basis, we may think of the Rogers-Szeg\"{o} polynomials as a
particularly interesting candidate to describe phase properties of the
deformed harmonic oscillator, with the additional advantage of automatically
allowing for the introduction of a parameter that can control the width of
the angle distribution function. In other words, we can see that the $\mu -$
parameter -- present in the Rogers-Szeg\"{o} polynomials variable -- may
stand for a squeezing parameter which controls the width of the angle
distribution function. %

\section{The Wigner function for the Rogers-Szeg\"{o} functions}

%
Although some proposals have been advanced in the past for defining the
Wigner function for the Abelian case of an angular momentum - angle degree
of freedom\cite{berry,mukunda,galetti,bizarro}, in the present case we must
draw our attention to the fact that the Rogers-Szeg\"{o} polynomials are
orthonormalized with respect to the Jacobi $\vartheta _{3}\left( \varphi
;q\right) $ measure function, in contrast to the cases discussed before when
that function is simply a constant.

Since $\vartheta _{3}\left( \varphi ;q\right) $ is an even function of $%
\varphi $ and guided by previous results \cite{berry,mukunda,galetti}, we
define the Weyl-Wigner mapped expression for an operator by taking the
Fourier transform namely{\bf ,} 
\begin{equation}
O\left( m,\theta \right) =\int_{-\pi }^{\pi }e^{im\widetilde{\theta }%
}\langle \theta -\frac{\widetilde{\theta }}{2}|\;\widehat{O}\;|\theta +\frac{%
\widetilde{\theta }}{2}\rangle \vartheta _{3}\left( \theta -\widetilde{%
\theta }/2;q\right) \frac{d\widetilde{\theta }}{2\pi }.  \label{IV1}
\end{equation}
This expression defines the quantum phase space representative of operator $%
\widehat{O}$. It must be noted that the choice for the $\vartheta _{3}\left(
\varphi ;q\right) $ argument could be $\theta +\frac{\widetilde{\theta }}{2}$
as well; the change would result in a change of sign in the $\widetilde{%
\theta }$ variable that leads, however, to the same final expression.

In this form, if we choose $\widehat{O}_{n}=|n\rangle \langle n|$, the
projector for the $n$-quanta harmonic oscillator (useful for writing a
density operator $\widehat{\rho }=\sum_{n=0}^{\infty }p_{n}|n\rangle \langle
n|$, $p_{n}$ being probabilities), we can get the Wigner function associated
to the Rogers-Szeg\"{o} polynomials. Therefore, the Wigner function
associated to one Rogers-Szeg\"{o} function 
\[
\langle \theta -\widetilde{\theta }/2|n\rangle =R_{n}\left( \theta -%
\widetilde{\theta }/2;q\right) 
\]
(and equivalently $\langle n|\theta +\frac{\widetilde{\theta }}{2}\rangle $
) will be, 
\[
O_{n}\left( m,\theta \right) =\int_{-\pi }^{\pi }e^{im\widetilde{\theta }%
}\langle \theta -\frac{\widetilde{\theta }}{2}|n\rangle \langle n|\theta +%
\frac{\widetilde{\theta }}{2}\rangle \vartheta _{3}\left( \theta -\widetilde{%
\theta }/2;q\right) \frac{d\widetilde{\theta }}{2\pi }. 
\]
Using the convergence of the series defining the $\vartheta _{3}\left(
\theta ;q\right) $ function, we get 
\[
O_{n}\left( m,\theta \right) =\sum_{t=-\infty }^{\infty }e^{-\mu
t^{2}+it\theta }\int_{-\pi }^{\pi }e^{im\widetilde{\theta }}\;e^{-it%
\widetilde{\theta }/2}R_{n}\left( \theta -\widetilde{\theta }/2;q\right)
R_{n}^{\ast }\left( \theta +\widetilde{\theta }/2;q\right) \frac{d\widetilde{%
\theta }}{2\pi }. 
\]
Now, using Eqs. (\ref{III8}) and (\ref{III1}) we get the expression 
\begin{eqnarray*}
O_{n}\left( m,\theta \right) &=&\frac{q^{n}}{\left( q,q\right) _{n}}%
\sum_{t=-\infty }^{\infty }e^{-\mu t^{2}+it\theta }\sum_{r,s=0}^{n}\left(
-1\right) ^{r+s}{%
{n \atopwithdelims[] r}%
}{%
{n \atopwithdelims[] s}%
}e^{\mu \left( r+s\right) }\;e^{i\theta \left( r-s\right) } \\
&&\times \int_{-\pi }^{\pi }e^{i\widetilde{\theta }\left( m-\left(
t+r+s\right) /2\right) }\frac{d\widetilde{\theta }}{2\pi }
\end{eqnarray*}
from which we obtain the Wigner function for the Rogers-Szeg\"{o}
polynomials 
\begin{eqnarray}
O_{n}\left( m,\theta \right) &=&\frac{q^{n}}{\left( q,q\right) _{n}}%
\sum_{t=-\infty }^{\infty }e^{-\mu t^{2}+it\theta }\sum_{r,s=0}^{n}\left(
-1\right) ^{r+s}{%
{n \atopwithdelims[] r}%
}{%
{n \atopwithdelims[] s}%
}e^{\mu \left( r+s\right) }\;e^{i\theta \left( r-s\right) }  \nonumber \\
&\times &\frac{\sin \left( m-\frac{t+r+s}{2}\right) \pi }{\left( m-\frac{%
t+r+s}{2}\right) \pi }.  \label{IV2}
\end{eqnarray}

Once we are given the Wigner function, we can extract the general form for
the probability distribution both for angle as for angular momentum as well.
To this end, we consider the general Wigner function, Eq. (\ref{IV2}).
First, integrating over the angle variable gives the angular momentum
distribution probability, viz. 
\begin{eqnarray*}
\Lambda ^{\left( n\right) }\left( m\right) &=&\int_{-\pi }^{\pi }\rho
^{\left( n\right) }\left( m,\theta \right) \frac{d\theta }{2\pi } \\
&=& \frac{q^{n}}{\left( q,q\right) _{n}}\sum_{r,s=0}^{n}\left( -1\right)
^{r+s} {%
{n  \atopwithdelims[] r}%
} {%
{n  \atopwithdelims[] s}%
} e^{\mu \left( r+s\right) }e^{-\mu \left( s-r\right) ^{2}}\;\frac{\sin
\left( m-s\right) \pi }{\left( m-s\right) \pi }.  \nonumber
\end{eqnarray*}
In order to carry out the summations in this expression, we must recall the
proof presented in the Appendix for the orthogonality of the
Rogers-Szeg\"{o} polynomials. Making use of expressions (A1, A2, A3), we see
that 
\[
\Lambda ^{\left( n\right) }\left( m\right) =\frac{q^{n}}{\left( q,q\right)
_{n}}\left( -1\right) ^{n}q^{\frac{n}{2}\left( n-1\right)
}\prod_{r=0}^{n-1}\left( 1-q^{r-n}\right) \frac{\sin \left( m-n\right) \pi }{%
\left( m-n\right) \pi }, 
\]
and, again from the orthogonality proof, we identify 
\[
\left( -1\right) ^{n}q^{\frac{n}{2}\left( n-1\right)
}\prod_{r=0}^{n-1}\left( 1-q^{r-n}\right) =\frac{\left( q,q\right) _{n}}{
q^{n}} 
\]
so that 
\begin{equation}
\Lambda ^{\left( n\right) }\left( m\right) =\frac{\sin \left( m-n\right) \pi 
}{\left( m-n\right) \pi }=\delta _{m,n}.  \label{IV3a}
\end{equation}

Therefore, the angular momentum distribution function only assumes the
discrete values associated to the polynomial indices, i.e., $m\geq 0$, thus
playing the role of an action variable.

On the other hand, by performing the summation over the angular momentum
variable we get the angle probability distribution, namely, 
\begin{eqnarray*}
\Omega ^{\left( n\right) }\left( \theta ,\mu \right) &=&\sum_{m=-\infty
}^{\infty }\rho ^{\left( n\right) }\left( m,\theta \right) =\frac{q^{n}}{%
\left( q,q\right) _{n}}\sum_{t=-\infty }^{\infty }e^{-\mu t^{2}+it\theta
}\;\sum_{r,s=0}^{n}\left( -1\right) ^{r+s} {%
{n  \atopwithdelims[] r}%
}{%
{n  \atopwithdelims[] s}%
} e^{\mu \left( r+s\right) }\;e^{i\theta \left( r-s\right) } \\
&&\times \sum_{m=-\infty }^{\infty }\frac{\sin \left( m-\frac{t+r+s}{2}%
\right) \pi }{\left( m-\frac{t+r+s}{2}\right) \pi }.
\end{eqnarray*}
Now, we have to observe that 
\[
\sum_{m=-\infty }^{\infty }\frac{\sin \left( m-\frac{t+r+s}{2}\right) \pi }{%
\left( m-\frac{t+r+s}{2}\right) \pi }=1 
\]
for $\left( t+r+s\right) /2$ integer or half-integer, therefore we get 
\[
\Omega ^{\left( n\right) }\left( \theta ,\mu \right) =\sum_{t=-\infty
}^{\infty }e^{-\mu t^{2}+it\theta }\left\{ \frac{q^{n}}{\left( q,q\right)
_{n}}\sum_{r,s=0}^{n}\left( -1\right) ^{r+s}{%
{n  \atopwithdelims[] r}%
} {%
{n  \atopwithdelims[] s}%
} e^{\mu \left( r+s\right) }\;e^{i\theta \left( r-s\right) } \right\}. 
\]
The curly bracket can be immediately identified as 
\[
\frac{q^{n}}{\left( q,q\right) _{n}}\sum_{r,s=0}^{n}\left( -1\right) ^{r+s}{%
{n  \atopwithdelims[] r}%
} {%
{n  \atopwithdelims[] s}%
} e^{\mu \left( r+s\right) }\;e^{i\theta \left( r-s\right) }=\left|
R_{n}\left( \theta ;\mu \right) \right| ^{2} 
\]
so that, finally, the angle distribution probability reads 
\begin{equation}
\Omega ^{\left( n\right) }\left( \theta ,\mu \right) =\sum_{t=-\infty
}^{\infty }e^{-\mu t^{2}+it\theta }\left| R_{n}\left( \theta ;\mu \right)
\right| ^{2}  \label{IV3b}
\end{equation}
as expected.

We can immediately verify that the Wigner function is normalized to unity by
just integrating Eq. (\ref{IV3b}) in the interval $-\pi \leq \theta <\pi $ ,
and recalling the orthogonalization procedure, or by summing expression ( 
\ref{IV3a}) over $m$ in the range $-\infty \leq m\leq \infty $.

As a first case of study it is now direct to particularize the Wigner
function to the lowest Rogers-Szeg\"{o} function, namely, let us consider $%
n=0$, the vacuum state projector. In this case 
\begin{equation}
O_n \left( m,\theta \right) =\sum_{t=-\infty }^{\infty }e^{-\mu
t^{2}+it\theta }\frac{\sin \left( m-\frac{t}{2}\right) \pi }{\left( m-\frac{t%
}{2}\right) \pi }  \label{IV3}
\end{equation}
that gives 
\[
\Lambda ^{\left( 0\right) }\left( m\right) =\delta _{m,n} 
\]
for the action probability distribution. On the other hand, the angle
probability distribution is 
\begin{equation}
\Omega ^{\left( 0\right) }\left( \theta ,\mu \right) =\vartheta _{3}\left(
\theta ;\mu \right) ,  \label{IVa}
\end{equation}
which is expected since for $n=0$, $R_{0}\left( \theta ;\mu \right) =1$.
This result coincides with the angle probability distribution as directly
obtained from the definition of the Rogers-Szeg\"{o} functions, Eq.( \ref
{III8}). This distribution is shown in Figure 1.

In the same form we can obtain the normalized angle probability distribution
for the second polynomial (projector $\widehat O _1$) that is written as 
\[
\Omega ^{\left( 1\right) }\left( \theta ,\mu \right) =\frac{e^{-2\mu }}{
1-e^{-2\mu }}\left( 1-2e^{\mu }\cos \theta +e^{2\mu }\right) \vartheta
_{3}\left( \theta ;\mu \right) , 
\]
and is depicted in Figure 2.

It is worth noticing that the angle probability distribution is $\mu$%
-dependent as expected, so that the width of $\Omega ^{\left( n\right)
}\left( \theta ,\mu \right) $ is governed by the free parameter $q$ (or
equivalently by $\mu $), which is the parameter of the deformed Heisenberg
algebra. %

\section{Summary and discussion}

%
In this contribution we have discussed the main properties of the
Rogers-Szeg\"{o} polynomials, and it was emphasized the fact that they are
in close connection with the eigenstates of the $q-$deformed harmonic
oscillator, as has been also pointed out in the Macfarlane realization.
Furthermore, knowing beforehand that these polynomials depend on an angular
variable defined on the circle, upon which they are orthonormalized with
respect to the Jacobi $\vartheta _{3}$- function, we propose that they can
be used as good functions to describe phase states. By making use of their
orthonormality requirements, we have also proposed a way to obtain the
Wigner function associated to them. Additionally, using the proper trace
operation techniques we have obtained the angle and action probability
distribution functions as a by-product of the Wigner function. It is seen
that the action variable only assumes values $m\geq 0$, and that the angle
distribution function is a well-behaved periodic function in the interval $%
-\pi \leq \theta <\pi $. In fact, this distribution function is immediately
identified as $\vartheta _{3}\left( \theta ;\mu \right) \left| R_{n}\left(
\theta ;\mu \right) \right| ^{2}$, as expected. %
\acknowledgments*{This work was supported by FAPESP under contract $\#$
00/15084-5. DG and SSM acknowledge partial financial support from CNPq
(Bras\'{i}lia)}. MR acknowledges financial support from CNPq. 
\appendix

\section{Carlitz orthogonality proof of the Rogers-Szeg\"{o} polynomials}

%
Let us first consider 
\begin{equation}
I_{mn}=\sum_{r=0}^{m}\sum_{s=0}^{n}\left( -1\right) ^{r+s} {%
{m  \atopwithdelims[] r}%
} {%
{n  \atopwithdelims[] s}%
} q^{\frac{r}{2}\left( r-1\right) }q^{\frac{s}{2}\left( s-1\right) }q^{-rs},
\label{a1}
\end{equation}
and let us show that $I_{mn}=q^{-n}\left( q;q\right) _{n}\delta _{n,m}$. To
this end we recall the result already presented in the text, Eq.(\ref{II5}) 
\[
\left( x;q\right) _{N}=\prod_{j=0}^{N-1}\left( 1-q^{j}x\right)
=\sum_{j=0}^{N}\left( -1\right) ^{j}{%
{N  \atopwithdelims[] j}%
} q^{\frac{1}{2}j\left( j-1\right) }x^{j} 
\]
and put it in (\ref{a1}) so that 
\begin{equation}
I_{mn}=\sum_{r=0}^{m}\left( -1\right) ^{r} {%
{m  \atopwithdelims[] r}%
} q^{\frac{r}{2}\left( r-1\right) }\prod_{s=0}^{n-1}\left( 1-q^{s-r}\right) .
\label{a2}
\end{equation}
Now, without any loss of generality, we can assume that $m\leq n$ (the
inverse could also be considered). There are two situations to be discussed.
First: For $m<n$, it is evident that the product on the rhs of (\ref{a2})
will vanish for all $r$ (the rhs is constituted of a sum of products. Each
summand has a product of terms where one of them will give $\left(
1-q^{r-r}\right) =0$, since, as $m<n$, $s$ will necessarily assume the value 
$r$). Therefore, the sum only have vanishing summands, since there will
always be a zero factor in the products.

Thus 
\begin{equation}
I_{mn}=0\;\;\;\;{\rm for{\ \ \ }}m<n{\rm {.}}  \label{a3}
\end{equation}
Second: For $m=n$ there will be only one term to be considered, namely $r=m$
, that will give 
\[
I_{nn}=\left( -1\right) ^{n}q^{\frac{n}{2}\left( n-1\right)
}\prod_{s=0}^{n-1}\left( 1-q^{s-n}\right) . 
\]
To calculate this expression, let us explicitly write the product 
\[
I_{nn}=\left( -1\right) ^{n}q^{\frac{n}{2}\left( n-1\right) }\left( 1-\frac{%
1 }{q^{n}}\right) \left( 1-\frac{1}{q^{n-1}}\right) \ldots \left( 1-\frac{1}{%
q} \right) 
\]
\[
=\frac{\left( 1-q^{n}\right) \left( 1-q^{n-1}\right) \ldots \left(
1-q\right) }{q^{n}} 
\]
which, upon identifying the numerator, gives 
\[
I_{nn}=\frac{\left( q;q\right) _{n}}{q^{n}}. 
\]
This contribution together with (\ref{a3}) gives the final result 
\[
I_{mn}=q^{-n}\left( q;q\right) _{n}\delta _{n,m}. 
\]
%

\medskip \newpage

{\bf Figure Captions}

\medskip \bigskip

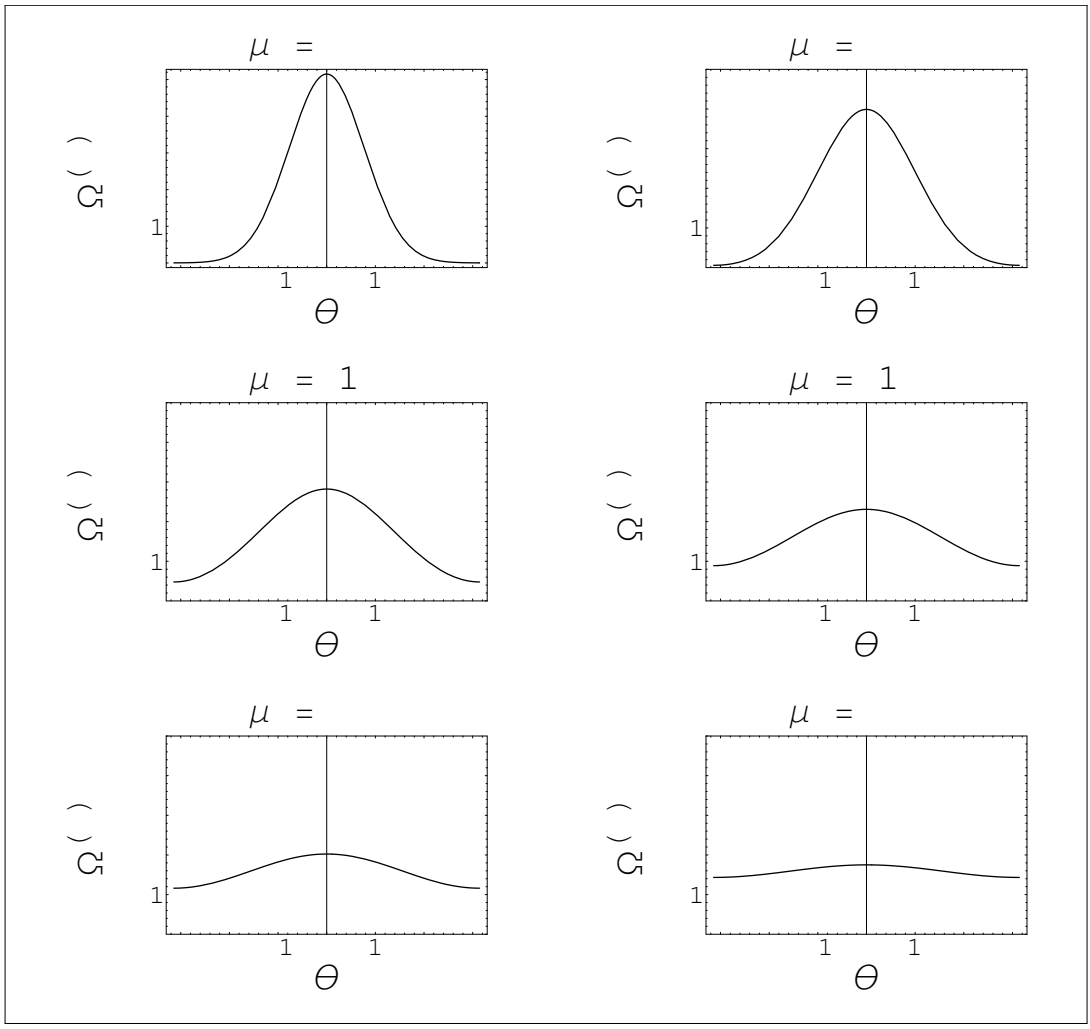
{\bf Figure 1}

Angle distribution functions associated to the first Rogers-Szeg\"{o}
polynomial obtained for some values of the parameter of the deformed
harmonic oscillator algebra.

\bigskip {\bf Figure 2}

Angle distribution functions associated to the second Rogers-Szeg\"{o}
polynomial obtained for some values of the parameter of the deformed
harmonic oscillator algebra.

%


\begin{references}
\bibitem{kulish}  Kulish P P and Reshetikhin N Y 1981 {\em Zap. Nauch, Sem.
LOMI\/} {\bf 101} 101-110 \newline
Kulish P P and Reshetikhin N Y 1983 {\em J. Sov. Math.} {\bf 23} 2435-2441 
\newline
Kulish P P and Sklyanin E K 1982 {\em Integrable Quantum Field Theories\/} ( 
{\em Lecture Notes in Physics\/} vol {\bf 151}), ed J Hietarinta (Berlin:
Springer) p 61-119

\bibitem{fadeev}  Faddeev L D 1984 {\em Recent Advances in Field Theory and
Statistical Mechanics (Proceedings of the Les Houches Summer School, Session
XXXIX, 1982)\/}, eds J B Zuber and R Stora (Amsterdam: North Holland) p
561-608

\bibitem{jimbo}  Jimbo M 1985 {\em Lett. Math. Phys}. {\bf 10 } 63-69

\bibitem{drinfeld}  Drinfel'd V G 1985 {\em Dokl. Akad. Nauk SSSR\/} {\bf 283%
} 1060-1064 \newline
Drinfel'd V G 1985 {\em Sov. Math. -- Doklady\/} {\bf 32} 254-258 \newline
Drinfel'd V G 1987 {\em Proc. Int. Congress Math. (Berkeley 1986)\/} vol 1,
ed A M Gleason (Providence: Am. Math. Soc.) p 798-820

\bibitem{manin}  Manin Y I 1987 {\em Ann. Inst. Fourier\/} {\bf 37}-(4)
191-205

\bibitem{iwata}  Iwata G 1951 {\em Prog. Theor. Phys.} {\bf 6} 524-528

\bibitem{arcoon}  Arik M, Coon D D and Lam Y M 1975 {\em J. Math. Phys.} 
{\bf 16} 1776-1779 \newline
Arik M and Coon D D 1976 {\em J. Math. Phys.} {\bf 17} 524-527

\bibitem{kurysh}  Kuryshkin V V 1976 {\em On some generalization of creation
and annihilation operators in quantum theory ($\mu $-quantization)\/}
(Moscow: VINITI) deposit N 3936--76 (in Russian)\newline
Kuryshkin V V 1980 {\em Ann. Fond. Louis de Broglie\/} {\bf 5} 111-125

\bibitem{cigler}  Cigler J 1979 {\em Monat. Math.} {\bf 88} 87-105

\bibitem{jan81}  Jannussis A, Brodimas G, Sourlas D and Zisis V 1981 {\em \
Lett. Nuovo Cim.} {\bf 30} 123-127 \newline
Brodimas G, Jannussis A, Sourlas D, Zisis V and Poulopoulos P 1981 {\em %
Lett. Nuovo Cim.} {\bf 31} 177-182

\bibitem{biedenharn}  Biedenharn L C 1989 {\em J. Phys. A: Math. Gen.} {\bf %
22} L873-878

\bibitem{macfarlane}  Macfarlane A J 1989 {\em J. Phys. A: Math. Gen.} {\bf %
22} 4581-4588

\bibitem{kudam}  Kulish P P and Damaskinsky E V 1990 {\em J. Phys. A: Math.
Gen.} {\bf 23} L415-L419

\bibitem{atsus}  Atakishiyev N M and Suslov S K 1991 {\em Theor. Math. Phys.}
{\bf \ 85} 1055 \newline
Atakishiyev N M and Suslov S K 1991 {\em Theor. Math. Phys.} {\bf 87} 442


\bibitem{sokat}  Solomon A I and Katriel J 1990 {\em J. Phys. A: Math. Gen.} 
{\bf 23} L1209-L1212 \newline
Solomon A I and Katriel J 1991 {\em J. Phys. A: Math. Gen.} {\bf 24}
2093-2105

\bibitem{fiv91}  Fivel D I 1991 {\em J. Phys. A: Math. Gen.} {\bf 24}
3575-3586

\bibitem{celeg91}  Celeghini E, Rasetti M and Vitiello G 1991 {\em Phys.
Rev. Lett.} {\bf 66} 2056-2059

\bibitem{chasri91}  Chaturvedi S and Srinivasan V 1991 {\em Phys. Rev.} A 
{\bf 44} 8020-8023 \newline
Chaturvedi S and Srinivasan V 1991 {\em Phys. Rev.} A {\bf 44} 8024-8026

\bibitem{jan91}  Jannussis A, Brodimas G and Mignani R 1991 {\em J. Phys. A:
Math. Gen.} {\bf 24} L775-L778 \newline
Brodimas G, Jannussis A and Mignani R 1992 {\em J. Phys. A: Math. Gen.} {\bf %
25} L329-L334

\bibitem{chiu92}  Chiu S-H, Gray R W and Nelson C A 1992 {\em Phys. Lett.} A 
{\bf 164} 237-242

\bibitem{ell92}  Ellinas D 1992 {\em Phys. Rev.} A {\bf 45} 3358-3361

\bibitem{manmen}  Man'ko V I and Mendes R V 1993 {\em Phys. Lett.} A {\bf 180%
} 39-42

\bibitem{alsol}  Solomon A I and Katriel J 1993 {\em J. Phys. A: Math. Gen.} 
{\bf 26} 5443-5447 \newline
McDermott R J and Solomon A I 1994 {\em J. Phys. A: Math. Gen.} {\bf 27}
L15-L19

\bibitem{camp94}  Campos R A 1994 {\em Phys. Lett.} A {\bf 184} 173-178

\bibitem{shan94}  Shanta P, Chaturvedi S, Srinivasan V and Jagannathan R
1994 {\em J. Phys. A: Math. Gen.} {\bf 27} 6433-6442

\bibitem{ck96}  Chung W-S and Klimyk A U 1996 {\em J. Math. Phys.} {\bf 37}
917-932

\bibitem{katq96}  Katriel J and Quesne C 1996 {\em J. Math. Phys.} {\bf 37}
1650-1661

\bibitem{park96}  Park S U 1996 {\em J. Phys. A: Math. Gen.} {\bf 29}
3683-3696

\bibitem{boda99}  Bonatsos D and Daskaloyannis C 1993 {\em Phys. Lett.} B 
{\bf 307} 100-105 \newline
Bonatsos D and Daskaloyannis C 1999 {\em Prog. Particle \& Nucl. Phys.} {\bf %
43} 537-618

\bibitem{isar02}  Isar A and Scheid W 2002 {\em Physica\/} A {\bf 310}
364-376 \newline
Isar A, Sandulescu A and Scheid W 2003 {\em Physica\/} A {\bf 322} 233-246

\bibitem{manko1}  Man'ko V I, Marmo G, Solimeno S and Zaccaria F 1993 {\em %
Int. J. Mod. Phys.} A {\bf 8} 3577-3597 \newline
Man'ko V I, Marmo G and Zaccaria F 1995 {\em Phys. Lett.} A {\bf 197} 95-99 
\newline
Manko V I, Marmo G, Sudarshan E C G and Zaccaria F 1997 {\em Phys. Scripta\/}
{\bf 55} 528-541 \newline
Man'ko V, Marmo G, Porzio A, Solimeno S and Zaccaria F 2000 {\em Phys. Rev.}
A {\bf 62} 053407

\bibitem{artbir}  Artoni M, Zang J and Birman J L 1993 {\em Phys. Rev.} A 
{\bf 47} 2555-2561

\bibitem{gruv}  Gruver J L 1999 {\em Phys. Lett.} A {\bf 254} 1-6

\bibitem{chak91}  Chakrabarti R and Jagannathan R 1991 {\em J. Phys. A:
Math. Gen.} {\bf 24} L711-L718

\bibitem{arik92}  Arik M, Demircan E, Turgut T, Ekinci L and Mungan M 1992 
{\em Z. Phys.} C {\bf 55} 89-95

\bibitem{chung}  Chung W-S, Chung K-S, Nam S-T and Um C-I 1993 {\em Phys.
Lett.} A {\bf 183} 363-370

\bibitem{3russos}  Borzov V V, Damaskinsky E V and Yegorov S B 1995 Some
remarks on the representations of the generalized deformed oscillator
algebra {\em Preprint\/} q-alg/9509022 \newline
Borzov V V, Damaskinsky E V and Yegorov S B 1997 {\em Zap. Nauch. Sem. LOMI\/%
} {\bf 245} 80-106 \newline
Borzov V V, Damaskinsky E V and Yegorov S B 2000 {\em J. Math. Sciences\/} 
{\bf 100} 2061

\bibitem{manko}  Man'ko V I, Marmo G, Solimeno S and Zaccaria F 1993 {\em %
Int. J. Mod. Phys.} A {\bf 8} 3577-3597 \newline
Man'ko V I, Marmo G and Zaccaria F 1995 {\em Phys. Lett.} A {\bf 197} 95-99 
\newline
Manko V I, Marmo G, Sudarshan E C G and Zaccaria F 1997 {\em Phys. Scripta\/}
{\bf 55} 528-541 \newline
Man'ko V, Marmo G, Porzio A, Solimeno S and Zaccaria F 2000 {\em Phys. Rev.}
A {\bf 62} 053407

\bibitem{micamdod}  Mizrahi S S, Camargo Lima J P and Dodonov V V 2004 {\em %
J. Phys. A: Math. Gen.} {\bf 37} 3707-3724 %
%

\bibitem{degroot}  S.R. de Groot and S.L. Suttorp, {\it Foundations of
Electrodynamics} (North Holland, Amsterdam, 1972).

\bibitem{carruthers}  P. Carruthers and F. Zachariasen, Rev. Mod Phys. {\bf %
55} (1983) 245.

\bibitem{balazs}  N.L. Balazs and B.K. Jennings, Phys. Rep. {\bf C104}
(1984) 347.

\bibitem{hillery}  M. Hillery, R.F. O 'Connel, M.O. Scully and E.P. Wigner,
Phys. Rep. {\bf C10} (1984).

\bibitem{kim}  Y.S. Kim and M.E. Noz, {\it Phase Space Picture of Quantum
Mechanics} (World Scientific, Singapore, 1991).

\bibitem{ozorio}  A.M. Oz\'{o}rio de Almeida, Phys. Rep. {\bf C295} (1998)
265.

\bibitem{berry}  M.V. Berry, Phil. Trans. Roy. Soc. {\bf 287} (1977) 237.

\bibitem{mukunda}  N. Mukunda, Am. J. Phys. {\bf 47} (1979) 182.

\bibitem{galetti}  D. Galetti and A.F.R. de Toledo Piza, Physica {\bf 149A}
(1988) 267.

\bibitem{bizarro}  J. Bizarro, Phys. Rev. {\bf A49} (1994) 3255.

\bibitem{szego}  G. Szeg\"{o}, {\it Orthogonal Polynomials}{\em \ } AMS
Colloquium Publications, vol. 23{\it , }Providence, USA, 1991).

\bibitem{szego2}  G. Szeg\"{o}, Sitzungsberichte Akad. Berlin (1926) 242.

\bibitem{carlitz1}  L. Carlitz, Ann. Math. Pura Applic. {\bf 41} (1956) 359.

\bibitem{carlitz}  L. Carlitz, Publicationes Mathematicae {\bf 5} (1958) 222.

\bibitem{andrews}  G.E. Andrews, {\it The Theory of Partitions} in
Enciclopedia of Mathematics and its Applications (Addison-Wesley Publishing
Company, 1976).


\bibitem{galetti2}  D. Galetti, Braz. Journ. of Physics. {\bf 33}, 148
(2003).

\bibitem{gasper}  G. Gasper and M. Rahman, {\it Basic Hypergeometric Series}
in Encyclopedia of Mathematics and its Applications (Cambridge University
Press, Cambridge, UK, 1997).

\bibitem{whittaker}  E.T. Whittaker and G.N. Watson, {\it A Course of} {\it %
Modern Analysis} (Cambridge University Press, USA, 1969).

\bibitem{bellman}  R. Bellman, {\it A Brief} {\it Introduction to Theta
Functions }(Holt, Rinehart and Winston, NY, 1961).

\bibitem{russo}  A.B. Klimov and S.M. Chumakov, Phys. Lett.{\bf A235 }(1997)
7.

\bibitem{gonz}  J. A. Gonz\'{a}lez and M.A. del Olmo, J. Phys. A: Math. Gen. 
{\bf 31}, 8841 (1998).

\bibitem{kowa}  K. Kowalski, J, Rembieli\'{n}ski and L. C. Papaloucas, J.
Phys. A: Math. Gen. {\bf 29}, 4149 (1996).

\bibitem{carr}  P. Carruthers and M. Martin Nieto, Rev. Mod. Phys. {\bf 40},
411 (1968).

\bibitem{suss}  L. Susskind and J. Glogower, Physics {\bf 1}, 49 (1964).

\bibitem{gama}  D. Galetti and M. A. Marchiolli, Ann. of Phys. {\bf 249},
454 (1996).

\bibitem{mehta}  M. L. Mehta, J. Math. Phys. {\bf 28}, 781 (1987).

\bibitem{feinsilver}  P. Feinsilver, Acta Appl. Math. {\bf 13,} 291 (1988).



\end{references}
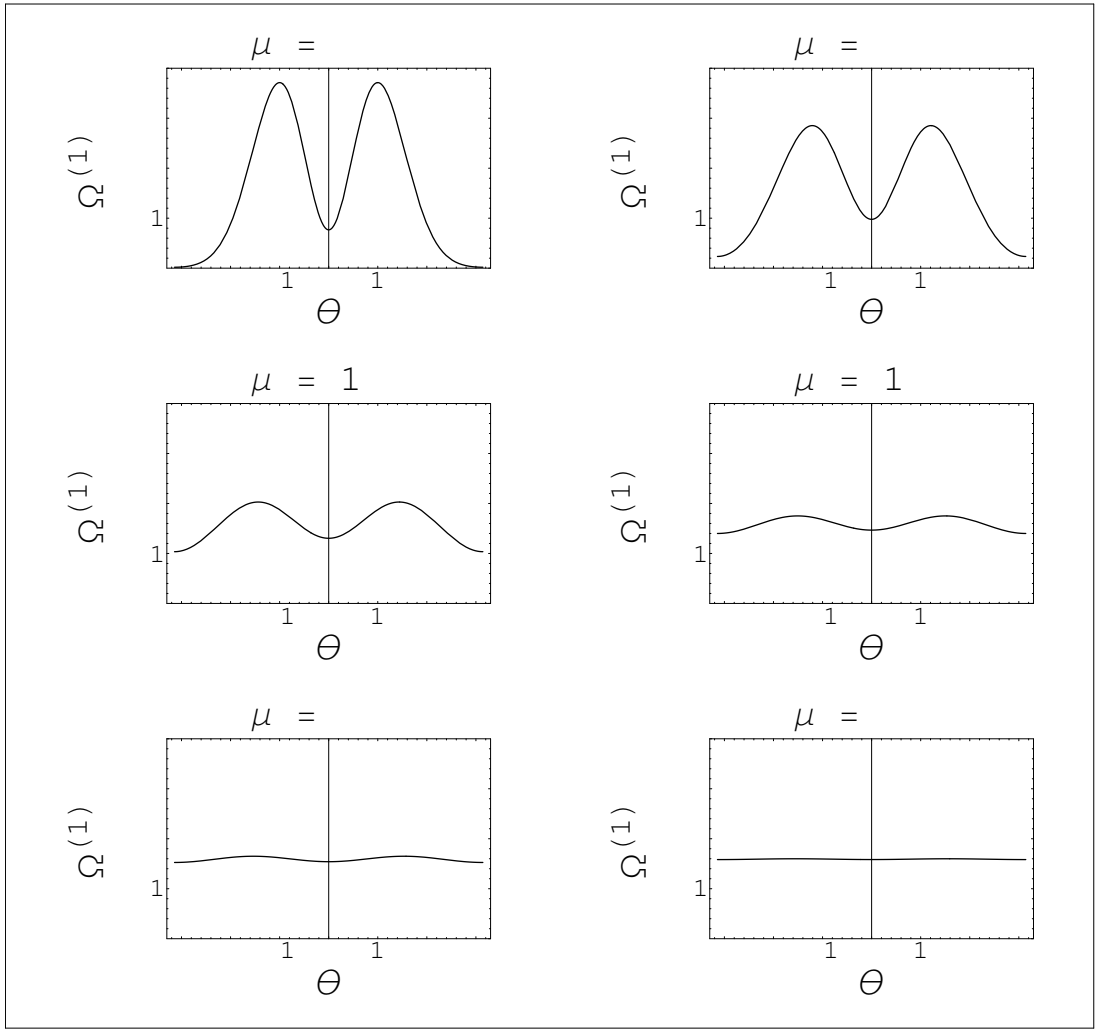
\end{document}